\documentclass{PoS}
\usepackage{graphicx}
\usepackage{epsfig}
\usepackage{wrapfig}
\usepackage[small,bf]{caption}
\usepackage{amsmath}
\usepackage{amssymb}
\usepackage{dsfont}
\usepackage{slashed}
\usepackage{multirow}
\usepackage{lscape}
\usepackage{wrapfig}

\newcommand{\Wuppertal}{Bergische Universit\"at Wuppertal, Theoretical Physics, 42119 Wuppertal, Germany.}
\newcommand{\Budapest}{E\"otv\"os University, Theoretical Physics, P\'azm\'any P. S 1/A, H-1117, Budapest, Hungary.}
\newcommand{\Regensburg}{Institute for Theoretical Physics, Universit\"at Regensburg, D-93040 Regensburg, Germany.}

\newcommand{\Julich}{J\"ulich Supercomputing Centre, Forschungszentrum J\"ulich, D-52425 J\"ulich, Germany.}

\newcommand{\be}{\begin{equation}}
\newcommand{\ee}{\end{equation}}

\newcommand{\Z}{\mathcal{Z}}

\title{The finite temperature QCD transition in external magnetic fields}
\ShortTitle{The finite temperature QCD transition in external magnetic fields}
\date{\today}
\author{
G.~S.~Bali$^1$, F.~Bruckmann$^1$, \speaker{G.~Endr\H{o}di}$^1$, Z.~Fodor$^2$, S.~D.~Katz$^3$, S.~Krieg$^{2,4}$, A.~Sch\"afer$^1$, K.~K.~Szab\'o$^2$
}

\footnotetext[1]{\Regensburg}
\footnotetext[2]{\Wuppertal}
\footnotetext[3]{\Budapest}
\footnotetext[4]{\Julich}

\abstract {
The effect of an external magnetic field on the finite temperature transition of QCD is studied. We measure thermodynamic observables including the quark condensates and susceptibilities and the strange quark number susceptibility. We generate configurations at various values of the quantized magnetic flux with $N_f=2+1$ flavors of stout smeared staggered quarks at physical quark masses. We perform the renormalization of our observables and approach the continuum limit with $N_t=6,8$ and $10$ lattices. We also check finite volume effects using various lattice volumes. Our main result is that the transition temperature significantly decreases with growing magnetic field, and that the transition remains an analytic crossover up to our largest external field $\sqrt{eB}\approx 1$ GeV.
}

\FullConference{The XXIX International Symposium on Lattice Field Theory \\
                July 11 - 16, 2011\\
                Squaw Valley, Lake Tahoe, California.}

\begin{document}

\section{Introduction}
\label{sec:intro}
\noindent
The response of the QCD vacuum to external (electro)magnetic fields is relevant for several physical situations, since very strong magnetic fields are thought to be present during the electroweak phase transition in the early universe~\cite{Vachaspati:1991nm}, in the interior of dense neutron stars called magnetars~\cite{Duncan:1992hi}, and in non-central heavy ion collisions~\cite{Skokov:2009qp}. In the latter case, although the external field has a very short lifetime (of the order of $1$ fm/$c$), the magnetic `impulse' coincides with the generation of the quark-gluon plasma and thus may have a significant effect on the properties of the transition.

The structure of the phase diagram of QCD in the magnetic field-temperature ($B$-$T$) plane has been studied extensively in the past years. Calculations have been carried out within various low energy effective models of QCD. Most of these models predict an increasing transition temperature $T_c$ and a strengthening of the transition with growing $B$, see, e.g.~\cite{Mizher:2010zb,Fraga:2008um,Gatto:2010pt,Kashiwa:2011js}. However, the opposite effect of a decreasing deconfinement transition temperature was predicted using chiral perturbation theory for two quark flavors~\cite{Agasian:2008tb}. Moreover, in most of these model calculations chiral symmetry breaking is enhanced with growing $B$ through an increase in the chiral condensate~\cite{Gusynin:1995nb,Nam:2011vn,Boomsma:2009yk}. On the other hand, it was also conjectured that the running of the strong coupling in the presence of magnetic fields may modify this magnetic catalysis, and even turn the effect around to make the dynamical mass decrease with $B$ in some regions~\cite{Miransky:2002rp}.

In a recent lattice simulation with $N_f=2$ flavors of staggered quarks~\cite{D'Elia:2010nq} the 
chiral condensate was observed to grow with the external field for any temperature $T$ in the transition region. The size of this effect was however found to be different for different values of $T$, resulting in an increase in both the pseudocritical temperature $T_c$ and the strength of the transition, being in qualitative agreement with most of the model predictions.
In this project our aim is to perform a similar lattice study, but with improved gauge and smeared fermionic actions and with $N_f=2+1$ flavors of quarks, at the physical pion mass, and extrapolate the results to the continuum limit.
We explore a wide temperature region around the zero-field pseudocritical temperature $T_c(B=0)$, for various values of the magnetic field, ranging up to $\sqrt{eB}\sim 1$ GeV, i.e. covering the region that is phenomenologically interesting for noncentral heavy ion collisions and for the evolution of the early universe. Our results are also published in a separate paper~\cite{magneticpaper}.

\section{Magnetic field on the lattice and observables}
\noindent
We consider a constant external magnetic field $\mathbf{B}=(0,0,B)$ in the $z$ direction, which can be realized by a vector potential $A_y=B x$ in the continuum.
On the lattice such a vector potential can be represented by complex phases $u_\nu(n)\in \mathrm{U}(1)$ that multiply the $U_\nu(n)\in \mathrm{SU}(3)$ links of the lattice~\cite{Martinelli:1982cb,Bernard:1982yu,Zhou:2002km,D'Elia:2010nq, Roberts:2010cz},
\be
\begin{split}
u_y(n) &= e^{i a^2 q B n_x}, \hspace*{3.05cm}u_x(N_x-1,n_y,n_z,n_t) = e^{-i a^2 q B N_x n_y}, \\
u_x(n) &= 1 \quad\quad\textmd{for } n_x\ne N_x-1, \hspace*{3cm}
u_\nu(n) =1 \quad\quad\textmd{for } \nu\not\in\{x,y\},
\end{split}
\label{eq:links2}
\ee
where the sites are labeled by integers $n=(n_x,n_y,n_z,n_t)$, with $n_\nu=0\ldots N_\nu-1$, and $N_\nu$ is the number of lattice points in the direction $\nu$. In this formulation we have periodic boundary conditions in all spatial directions and the magnetic flux going through any plaquette in the $x-y$ plane is constant.

It is well known that in a finite box with periodic boundary conditions the magnetic flux is quantized in terms of the area of the plane orthogonal to the external field~\cite{'tHooft:1979uj,AlHashimi:2008hr}. On the lattice this leads to the quantization condition,
\be
a^2 qB = \frac{2\pi N_b}{N_x N_y}, \quad\quad\quad N_b \in \mathds{Z},
\label{eq:quantcont}
\ee
where $q$ is the smallest charge in the system -- in our case the down quark charge $q=q_d$. We note that the lattice magnetic field is periodic in $N_b$ with a period of $N_xN_y$. In order to have an unambiguous implementation we constrain the flux such that $0\le N_b< N_xN_y/4$. At larger values of $N_b$ the periodicity is expected to introduce saturation effects, like it was observed in~\cite{D'Elia:2011zu,Bruckmann:2011zx}. 

We derive our observables from the staggered partition function with three flavors $u,d$ and $s$. The quark flavors are treated separately since their charges/masses are different: $q_u=-2q_d=-2q_s$, and we assume $m_u=m_d\ne m_s$. The dependence on the external field enters the partition function only through the determinants, in the form $q_fB$. Since we are concerned with a constant external field we do not include any dynamics for the $\mathrm{U}(1)$ field introduced above. However our approach is `dynamical' (i.e. not quenched) in the sense that the magnetic field is taken into account both in the configuration production and in the measurements.

Our observables include the chiral condensates and susceptibilities for the light flavors $f=u,d$, and the strange quark number susceptibility,
\be
\bar{\psi}\psi_f \equiv \frac{T}{V}\frac{\partial \log \Z}{\partial m_f}, \quad\quad\quad\quad \chi_f \equiv \frac{\partial \bar{\psi}\psi_f}{\partial m_f}, \quad\quad\quad\quad c_2^s \equiv \frac{T}{V}\frac{1}{T^2}\frac{\partial^2 \log \Z}{\partial \mu_s^2},
\label{eq:defc2s}
\ee
where we defined the spatial volume of the system as $V= (N_sa)^3$ with $N_s\equiv N_x=N_y=N_z$.
The condensate will be denoted in the following by the first letter of the flavor name, e.g. $\bar u u$. 

In order to approach the continuum limit, the renormalization of these observables has to be carried out. We remark that the divergences to be canceled are independent of both $T$ and $B$~\cite{Salam1975203,magneticpaper}.
Therefore we can eliminate the additive divergences by subtracting the $T=0$, $B=0$ contribution. Then we multiply by an appropriate power of the bare quark mass to cancel the multiplicative divergences~\cite{Endrodi:2011gv},
\be
\bar{\psi}\psi_f^r (B,T) = m_f \left [\bar{\psi}\psi_f(B,T) - \bar{\psi}\psi_f(0,0) \right] \frac{1}{m_\pi^4}, \quad
\chi_f^r (B,T) = m_f^2 \left [\chi_f(B,T) - \chi_f(0,0) \right] \frac{1}{m_\pi^4}.
\ee
This procedure leads to a renormalized condensate that, for $B=0$, is zero at $T=0$ and approaches a negative value as $T$ is increased.
Considering the strange quark number susceptibility, $c_2^s$ needs no renormalization (neither at $B=0$ nor at $B\ne 0$) since it is connected to a conserved current.

In this study we use the tree-level improved Symanzik gauge action and stout smeared staggered fermions; details about the action can be found in~\cite{Aoki:2005vt}. We generate lattice configurations both at $T=0$ and $T>0$ with an exact RHMC algorithm, for various values of the gauge coupling and the magnetic flux. For our finite temperature runs we have lattice configurations with $N_t=6,8$ and $10$. Finite volume effects are studied on the $N_t=6$ ensemble using sets of $N_s=16$, $24$ and $32$ lattices. The masses of the up, down and strange quarks are set to their physical values along the line of constant physics (LCP) by fixing the ratios $f_K/m_\pi$ and $f_K/m_K$ to their experimental values. The lattice spacing is determined by $f_K$. Details of the determination of the LCP and the lattice scale can be found in, e.g.~\cite{Borsanyi:2010cj}. Since the finite external field does not affect the lattice scale~\cite{magneticpaper}, we use the lattice spacing measurements at $T=0$ and $B=0$ to set the scale also at $T \ne 0$ and $B\ne 0$. 

\section{Chiral condensate at finite $T$ and $B$}
\noindent

\begin{wrapfigure}{r}{6.0cm}
\centering
\vspace*{-0.8cm}
\hspace*{-0.2cm}\includegraphics*[width=6.4cm]{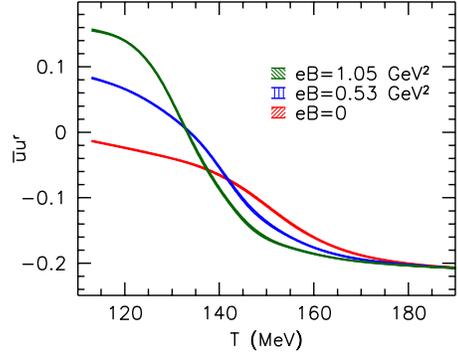}
\vspace*{-0.8cm}
\caption{The renormalized chiral condensate as a function of $T$ for three different values of $B$ for the $N_t=6$ lattices. The transition temperature shifts towards the left.}
\label{fig:pbp_T}
\end{wrapfigure}

In figure~\ref{fig:pbp_T} we plot the renormalized chiral condensate as a function of the temperature for three values of $B$. The transition temperature apparently decreases with increasing $B$, thereby contradicting a vast number of model calculations, see the summary given in the introduction. Furthermore this observation also disagrees with the lattice result of~\cite{D'Elia:2010nq}. First of all, to check our simulation code we reproduced the results of~\cite{D'Elia:2010nq} with an identical setting (i.e. same lattice sizes and quark masses, no smearing and no improvement in the action) at a couple of temperatures. Since we find a perfect agreement, we conclude that we are left with three possible reasons for the discrepancy. First, the lattice spacing of~\cite{D'Elia:2010nq} is larger, $a\approx 0.3$ fm, and also an unimproved action is used, so lattice discretization errors may be significant. Second, the present study uses $N_f=2+1$ flavors as opposed to the $N_f=2$ of~\cite{D'Elia:2010nq}, and the pseudocritical temperature is known to depend on the number of flavors~\cite{Karsch:2000kv}, which may also introduce systematic differences in the dependence on the external field. Third, the quark masses of~\cite{D'Elia:2010nq} are larger than in the present study, which can also cause drastic changes in thermodynamics -- for example the nature of the transition at $B=0$ depends very strongly (and non-monotonically) on the quark masses.

On closer inspection, the differences between our results and those of~\cite{D'Elia:2010nq} can actually be traced back to the behavior of the chiral condensate as a function of $B$ and $T$. While the authors of~\cite{D'Elia:2010nq} observed that at any temperature the condensate increases with $B$, we find that this dependence is more complex and in the transition region the condensate decreases with growing $B$. As already mentioned in the introduction, the possibility of such a decrease in the condensate with $B$ was also raised in low energy model calculations~\cite{Miransky:2002rp}. We find that for three strange quarks this complex dependence turns into a monotically increasing behavior for all temperatures~\cite{magneticpaper}, which indicates that the response of the chiral condensate to the external field is very strongly influenced by the quark masses.
We summarize our findings as a) the dependence of the condensate on the external field is non-monotonic and varies strongly with temperature, and b) as a result the pseudocritical temperature shifts to lower values at large $B$ as compared to the $B=0$ case. The latter observation is supported by a similar $T_c(B)$ dependence deduced from the chiral susceptibility, the Polyakov loop and the strange susceptibility~\cite{magneticpaper}. 

\section{Transition strength and transition temperature}
\label{sec:finitesize}
\noindent
First we study the strength of the transition as a function of the external field.
At $B=0$ the transition is known to be a broad crossover~\cite{Aoki:2006we}, where the approximate order parameters like the chiral condensate and the Polyakov loop change smoothly with the temperature, and no finite volume scaling is visible in the observables. 

\begin{figure}[h!]
\centering
\vspace*{-0.3cm}
\mbox{
\includegraphics*[width=5.4cm]{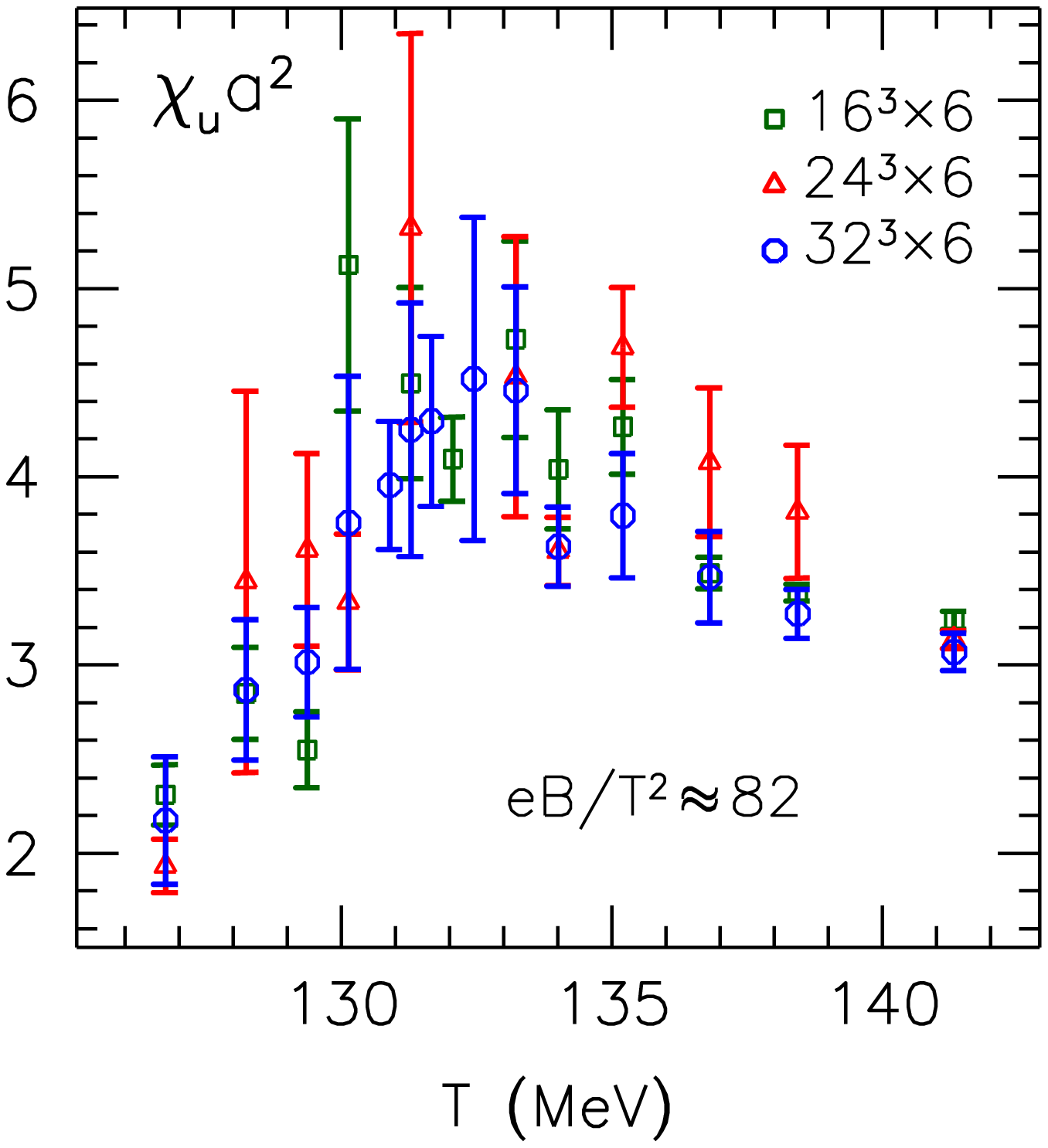}\hspace*{0.5cm}
\includegraphics*[width=7.2cm]{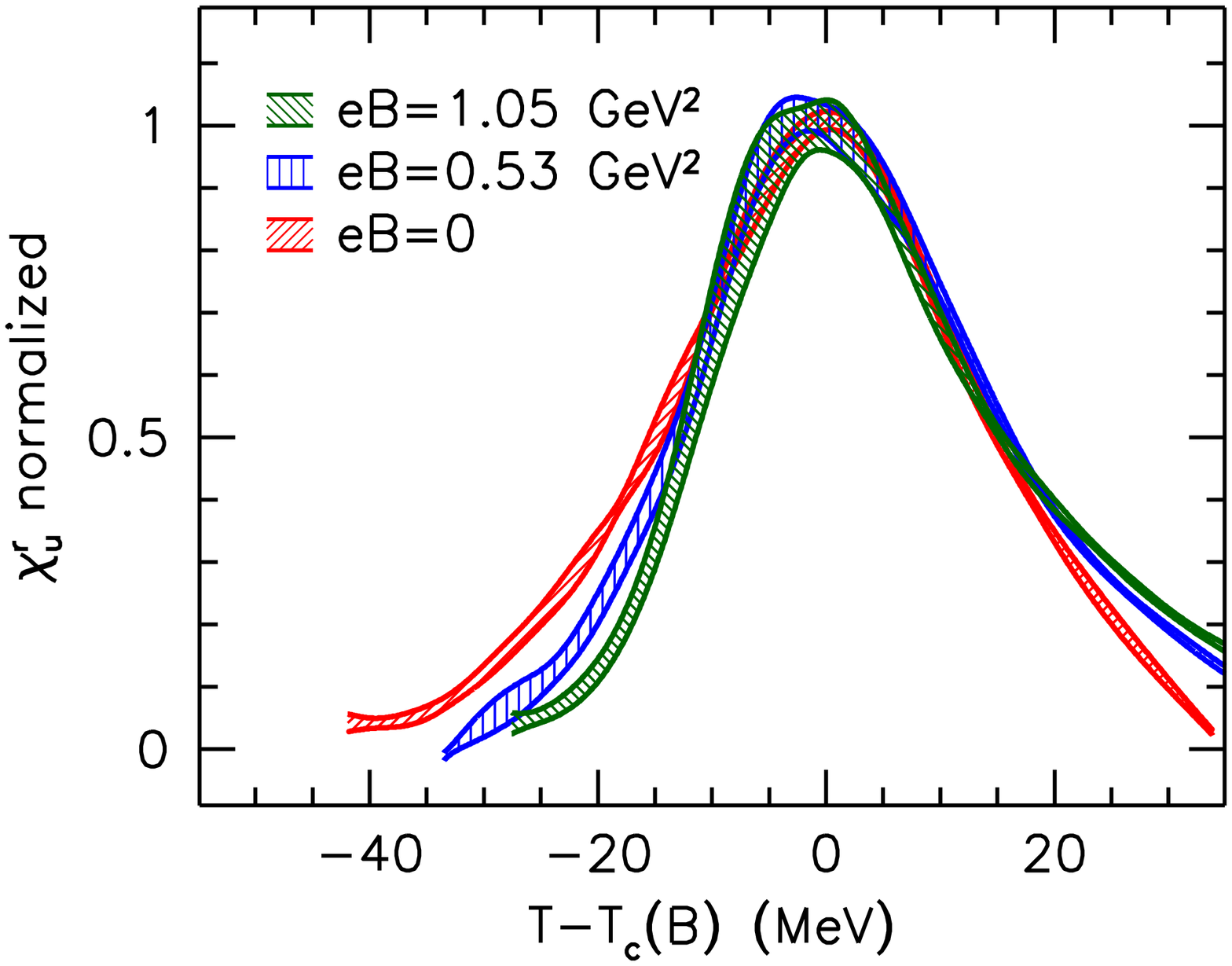}
}
\vspace*{-0.1cm}
\caption{The chiral susceptibility as a function of $T$ measured on our $N_t=6$ lattices for different spatial volumes (left panel). Relative changes in the $T$-dependence of $\chi_u^r$ as measured on $N_t=6$ lattices (right panel).}
\label{fig:voldep}
\end{figure}

To properly determine the nature of the transition we search for finite volume scaling in our observables.
To this end we perform simulations at our largest magnetic field on the $N_t=6$ lattices with $N_s=16,24$ and $32$. The largest lattice in the transition region corresponds to a box of linear size $\sim 7$ fm. 
Here we keep $eB/T^2$ fixed (and not $B$ itself) as we are only interested in differences between the various volumes. In the left panel of figure~\ref{fig:voldep} the results for the chiral susceptibility are shown as a function of the temperature for $eB/T^2\approx82$. 
The figure shows that our $N_s=16$ results agree within statistical errors with the $N_s=24$ and $N_s=32$ data, indicating that finite size errors are small, compared to statistical errors. This independence on the volume also implies that the transition at this high magnetic field is still an analytic crossover. 

To further study how the strength of the transition changes we investigate the width of the renormalized chiral susceptibility. In the right panel of figure~\ref{fig:voldep} we plot the susceptibility divided by its maximum value 
as a function of $T-T_c(B)$ for three different values of the magnetic field for the $N_t=6$ lattices. 
We find that the width of the peak is only mildly affected by the magnetic field. In particular, the width of the peak at half maximum decreases from $\sim 30(3)$ MeV to $\sim 25(3)$ MeV as the external field is increased from zero to $eB=1.05 \textmd{ GeV}^2$. 
We find a very similar behavior on the $N_t=8$ and $10$ lattices.
From this analysis our final conclusions are that the width of the transition decreases only mildly with increasing magnetic field, and as the finite size scaling analysis has shown, the transition remains an analytic crossover at least up to $\sqrt{eB}\sim 1$ GeV.

\begin{figure}[h!]
\centering
\vspace*{-0.5cm}
\mbox{
\includegraphics*[width=7.0cm]{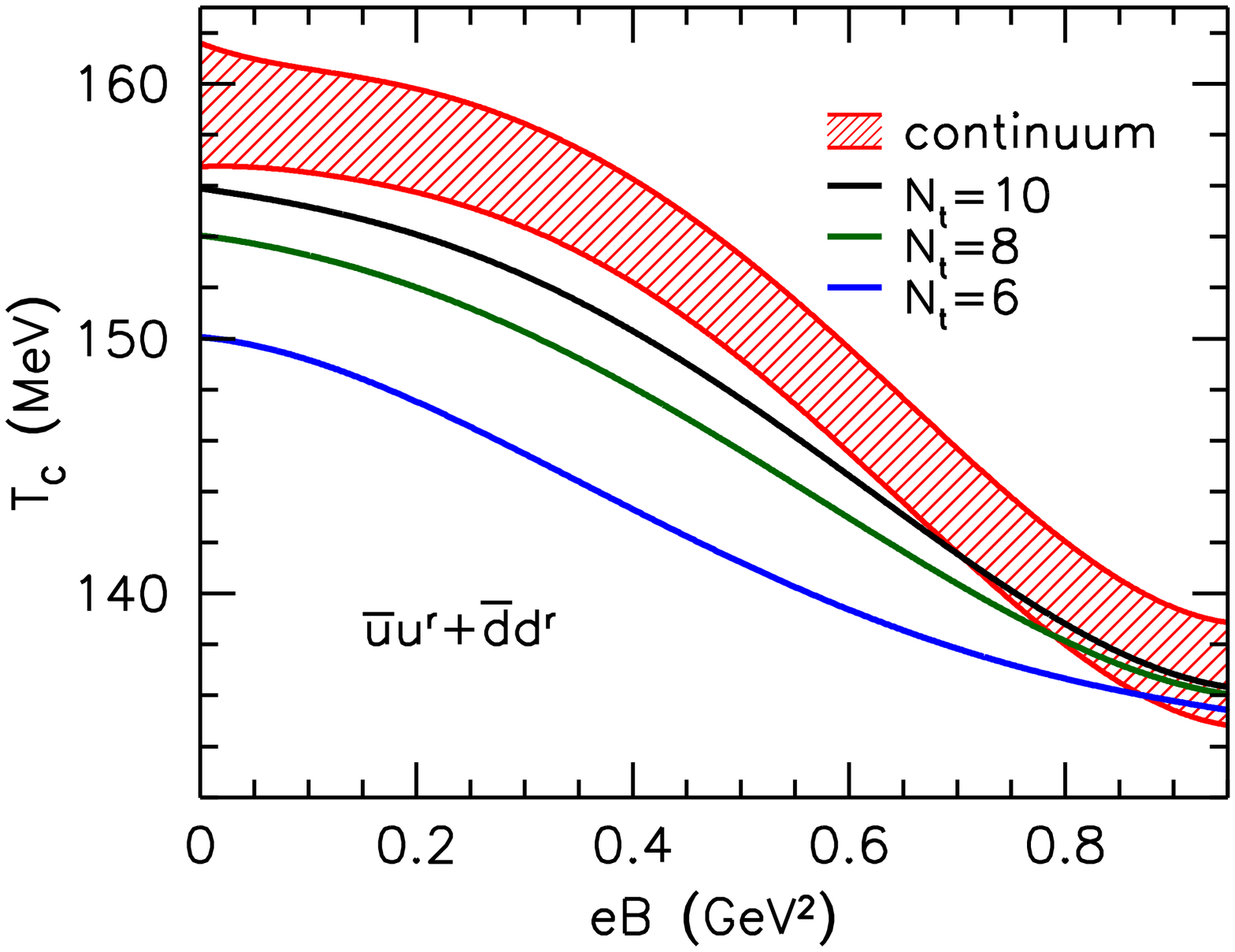}\quad
\includegraphics*[width=7.0cm]{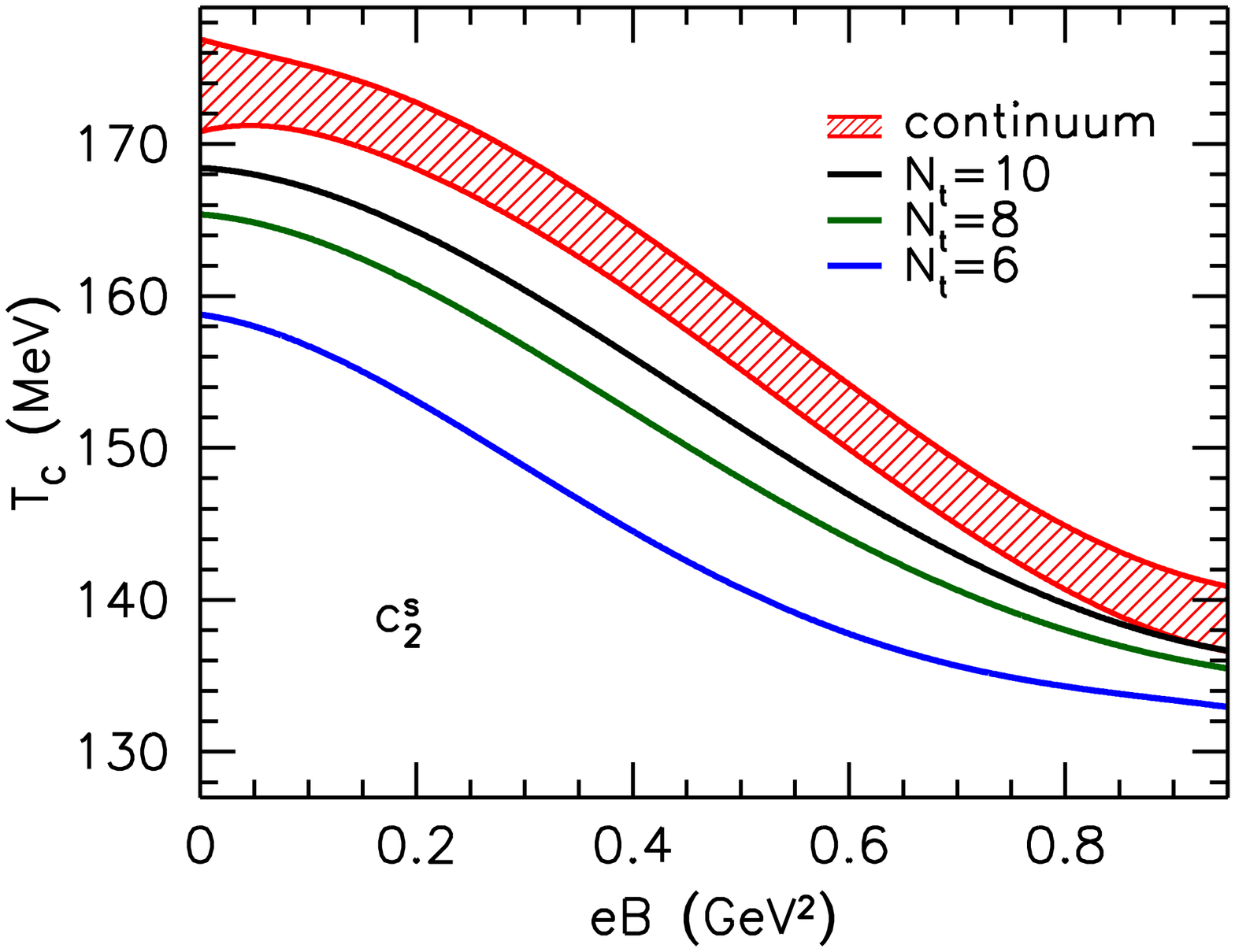} }
\vspace*{-0.2cm}
\caption{The phase diagram of QCD in the $B-T$ plane, determined from the renormalized chiral condensate $\bar u u^r + \bar d d^r$ (left panel) and the strange quark number susceptibility $c_2^s$ (right panel).}
\label{fig:phasediag_cont}
\end{figure}

Next we study the observables as functions of the temperature to determine the pseudocritical temperature.
We search for the inflection point of the renormalized chiral condensate $\bar u u^r + \bar d d^r$ and the strange quark number susceptibility $c_2^s$. To carry out the continuum extrapolation, we fit the results for $T_c(B)$ for all three lattice spacings ($N_t=6,8$ and $10$) together with an $N_t$-dependent polynomial function of order four of the form $T_c(B,N_t) = \sum_{i=0}^4(a_i + b_i N_t^{-2}) B^i$.
This ensures the scaling of the final results with $N_t^{-2}\sim a^2$. We obtain $\chi^2/{\rm dof.}\approx 0.5 \ldots 1.2$ indicating good fit qualities. The results are plotted for the condensate and the strange susceptibility in figure~\ref{fig:phasediag_cont}. 

As is clearly visible in the figure, both observables show that the pseudocritical temperature {\it decreases} with growing external field $B$. We mention that the chiral susceptibility gives a similar phase diagram, and that preliminary results about the Polyakov loop also indicate a decrease in $T_c(B)$~\cite{magneticpaper}.

\section{Summary}
\noindent
In this paper we studied the finite temperature transition of QCD in the presence of external (electro)magnetic fields via lattice simulations at physical quark masses. The extrapolation to the continuum limit is carried out, and finite size effects are under control. The results are relevant for the description of both the evolution of the early universe and of noncentral heavy ion collisions. 

We obtained the phase diagram of QCD in the $B-T$ plane in the phenomenologically interesting region of $0\le eB \lesssim 1 \textmd{ GeV}^2$. 
Performing a finite volume scaling study we found that the transition remains an analytic crossover up to our largest magnetic fields, with the transition width decreasing only mildly.
This rules out the existence of a critical endpoint in the $B-T$ phase diagram below $eB=1 \textmd{ GeV}^2$.

Moreover, our results indicate that the transition temperature significantly {\it decreases} with increasing $B$. This result contradicts several model calculations present in the literature which predict an increase in $T_c$ as $B$ grows (see the summary in section~\ref{sec:intro}). 
By comparing our magnetic fields to the maximal fields that may be produced in noncentral heavy ion collisions we conclude that the decrease in $T_c$ is negligible for RHIC and may be up to $5-10$ MeV for the LHC. Moreover, the effect grows with the magnetic field, exceeding 20$\%$ for $c_2^s$ at $\sqrt{eB}=1$ GeV. This may have a significant impact on the description of the QCD transition during the evolution of the early universe.

\begin{acknowledgments}
\noindent
This work has been supported by DFG grants SFB-TR 55, FO 502/1-2 and BR 2872/4-2, the EU grants (FP7/2007-2013)/ERC no 208740 and PITN-GA-2009-238353 (ITN STRONGnet). Computations were carried out on the GPU cluster at the E\"otv\"os University in Budapest and on the Bluegene/P at FZ J\"ulich. We thank Ferenc Niedermayer for useful discussions, interesting ideas and for careful reading of the manuscript. G.E. would like to thank Massimo D'Elia, Swagato Mukherjee, D\'aniel N\'ogr\'adi, Tam\'as Kov\'acs and Igor Shovkovy for useful discussions. 
\end{acknowledgments}

\bibliographystyle{JHEP}
\bibliography{magn_pos}
\end{document}